%% file: GRSL_GSSTV_arXiv.tex
\documentclass[journal]{IEEEtran}
\def\x{{\mathbf x}}

\usepackage{amssymb}
\usepackage{amsthm}
\usepackage{booktabs}

\usepackage{cite}

\ifCLASSINFOpdf
  \usepackage[pdftex]{graphicx}
\else
  \usepackage[dvips]{graphicx}
\fi
\usepackage{amsmath}
\usepackage{rotating}
\usepackage{multirow}
\usepackage{url}
\usepackage{color}

\usepackage{algorithm}
\usepackage{algorithmic}
\begin{document}
%
\title{Graph Spatio-Spectral Total Variation Model\\ for Hyperspectral Image Denoising}
%
%
%

\author{Shingo~Takemoto,
        Kazuki~Naganuma,~\IEEEmembership{Student~Member,~IEEE,}
        and~Shunsuke~Ono,~\IEEEmembership{Member,~IEEE,}
\thanks{Manuscript received XXX, XXX; revised XXX XXX, XXX.}%
\thanks{This work was supported in part by JST PRESTO under Grant JPMJPR21C4, and in part by JSPS KAKENHI under Grant 22H03610, 22H00512, 20H02145, and 18H05413.}
\thanks{S. Takemoto is with the Department of Computer Science, Tokyo Institute of Technology, Yokohama, 226-8503, Japan (e-mail: takemoto.s.af@m.titech.ac.jp).}
\thanks{K. Naganuma is with the Department of Computer Science, Tokyo Institute of Technology, Yokohama, 226-8503, Japan (e-mail: naganuma.k.aa@m.titech.ac.jp).}
\thanks{S. Ono is with the Department of Computer Science, Tokyo Institute of Technology, Yokohama, 226-8503, Japan (e-mail: ono@c.titech.ac.jp).}}

\maketitle

\begin{abstract}
The spatio-spectral total variation (SSTV) model has been widely used as an effective regularization of hyperspectral images (HSI) for various applications such as mixed noise removal. 
However, since SSTV computes local spatial differences uniformly, it is difficult to remove noise while preserving complex spatial structures with fine edges and textures, especially in situations of high noise intensity. 
To solve this problem, we propose a new TV-type regularization called Graph-SSTV (GSSTV), which generates a graph explicitly reflecting the spatial structure of the target HSI from noisy HSIs and incorporates a weighted spatial difference operator designed based on this graph. 
Furthermore, we formulate the mixed noise removal problem as a convex optimization problem involving GSSTV and develop an efficient algorithm based on the primal-dual splitting method to solve this problem. 
Finally, we demonstrate the effectiveness of GSSTV compared with existing HSI regularization models through experiments on mixed noise removal.
The source code will be available at \url{https://www.mdi.c.titech.ac.jp/publications/gsstv}.
\end{abstract}


%
\IEEEpeerreviewmaketitle

\section{Introduction}
\label{sec:intro}

\IEEEPARstart{H}{yperspectral} images (HSIs) have rich spatial and spectral information and offer many potential applications in a wide range of fields, such as agriculture, mineralogy, astronomy, and biotechnology \cite{borengasser2007hyperspectral,grahn2007techniques,thenkabail2016hyperspectral}. In the process of acquiring and transmitting HSIs, various types of noise, such as thermal noise, quantization noise and shot noise, are inevitable. Such noise significantly degrades the performance of subsequent processing, including unmixing~\cite{HSI_unmixing_review,unmixing2014} and classification~\cite{ghamisi2017advanced,audebert2019deep, Classification_DR, SAHDA}, and so denoising has become an essential task for hyperspectral imaging \cite{rasti2018noise,HSI_review_Ghamisi2017}.

For HSI denoising tasks, the Spatio-Spectral Total Variation (SSTV) model~\cite{SSTV} is known as a powerful regularization approach that adequately captures the spectral structure of HSIs, and has been widely used in state-of-the-art HSI denoising methods~\cite{LRMRSSTV,SSTV_LRTD,GLSSTV,SSTV_Wang2021}.
On the other hand, as spatial regularization, SSTV simply evaluates the neighboring differences along the vertical and horizontal directions, so there is still room for improvement to achieve denoising that preserves the complex spatial structure of 
HSIs with mixed edges and textures.

A promising approach to fully capture complex spatial structures is to construct a graph that reflects the spatial structure and design regularization models via the graph. In fact,  graph total variation and its generalized models \cite{graph_image_denoising_2008,couprie2013dual,ono2015total,chen2015signal,berger2017graph} and a patch-based graph regularization model \cite{graph_image_denoising_Gene2017} constructed in such a way have achieved high-quality denoising that preserves detailed spatial structure in applications to natural image and depth map processing.

Inspired by these graph-based approaches, this paper proposes a mixed noise removal method for HSI using a newly formulated \textit{Graph Spatio-Spectral Total Variation} model (GSSTV). Our main contribution is twofold. The first is the new formulation of a regularization, namely GSSTV, that can adequately capture both the spatial and spectral characteristics of the target HSI. GSSTV consists of integrating SSTV with a weighted spatial difference operator defined using a graph that explicitly reflects the spatial structure of the target HSI (called the spatial graph). In doing so, we also propose a method to robustly construct a spatial graph from a given noisy observed HSI. The second is the formulation of the mixed noise removal problem as a constrained convex optimization problem involving GSSTV and the construction of an algorithm for solving this problem based on a primal-dual splitting method \cite{PDS_pock}. In the proposed formulation, terms other than the GSSTV, such as a data fidelity term, are written as convex constraints rather than as part of the objective function, which greatly simplifies parameter adjustment. In addition, the proposed algorithm does not require matrix inversion, singular value decomposition, etc., and thus can compute the optimal solution (= the restored HSI) efficiently. Finally, we demonstrate the effectiveness of the proposed method by comparing it with several existing HSI regularization models, including SSTV, through experiments of mixed noise removal.

\section{Preliminaries}
\label{sec:preliminaries}

\subsection{Notations}
\label{subsec:notations}
Throughout this paper, we denote vectors and matrices by the boldface lowercase letters (e.g., $\mathbf{x}$) and boldface capital letters (e.g., $\mathbf{X}$), respectively. For a vectorized matrix data $\mathbf{x}\in\mathbb{R}^{N_{1}N_{2}}$, the value of the location $(i,j)$ is denoted by $[\mathbf{x}]_{i,j}$. For cube data $\mathcal{X}\in\mathbb{R}^{N_{1}\times N_{2} \times N_{3}}$, let the vectorized $k$th frontal slices of $\mathcal{X}$ be $\mathrm{vec}(\mathbf{X}_{k})$, then its vectorized form of $\mathcal{X}$ is defined as $\mathbf{x}=\begin{pmatrix}\mathrm{vec}(\mathbf{X}_{1})^{\top} & \cdots & \mathrm{vec}(\mathbf{X}_{N_{3}})^{\top} \end{pmatrix}^{\top} \in \mathbb{R}^{N_{1}N_{2}N_{3}}$. We denotes the number $N_{1}N_{2}N_{3}$ of cube data elements to $N$. Similar to the matrix case, the value of the vectorized cube data $\mathbf{x}$ of the location $(i,j,k)$ is denoted by $[\mathbf{x}]_{i,j,k}$. 
The $\ell_1$-norm, the $\ell_2$-norm, and the signum function are denoted by $\| \cdot \|_1$, $\| \cdot \|_2$, and $\mathrm{sgn}(\cdot)$, respectively.

\subsection{Spatio-Spectral Total Variation (SSTV) \textup{\cite{SSTV}}}
\label{subsec:SSTV}

For the vecterized cube data $\mathbf{u} \in \mathbb{R}^{N}$, SSTV is defined as
\begin{equation}
	\label{eq:SSTV}
	\mathrm{SSTV}(\mathbf{u}) := \| \mathbf{D}_v \mathbf{D}_b \mathbf{u} \|_1 + \| \mathbf{D}_h \mathbf{D}_b \mathbf{u} \|_1,
\end{equation}
where $\mathbf{D}_h \in \mathbb{R}^{N\times N}$, $\mathbf{D}_v \in \mathbb{R}^{N \times N}$, and $\mathbf{D}_b \in \mathbb{R}^{N \times N}$ are the forward difference operators in the horizontal, vertical, and spectral directions, and the boundary condition is the Neumann boundary condition. 
SSTV does not simply evaluate the spatial differences but also evaluates the spatial differences multiplying by the spectral difference operator.
Therefore, SSTV can remove noise in HSI well while retaining the consistent spatio-spectral structure.

\subsection{Graph-Based Weighted Spatial Difference Operator \textup{\cite{GSIP}}}
\label{subsec:GTV}
For a given grayscale guide image denoted by $\mathbf{x}\in \mathbb{R}^{N_1 N_2}$, we consider to construct a weighted graph $\mathcal{G}(\mathbf{x}, \mathcal{E}, \mathbf{W})$ with edges $e_{p,q} \in \mathcal{E}$ ($p$ and $q$ are indices of pixels ($1 \leq p<q \leq N_{1}N_{2}$)).
The weight matrix $\mathbf{W}\in\mathbb{R}^{|\mathcal{E}|\times |\mathcal{E}|}$ ($|\mathcal{E}|$ is the number of edges) is a diagonal matrix whose entries are the weights $w_{e_{p,q}}\in(0,1]$ assigned to the edges $e_{p,q}$, defined as
\begin{equation}
    \label{eq:graph_weight}
	w_{e_{p,q}} := \exp{(-\sigma_l^{-1}\| \mathbf{l}_p - \mathbf{l}_q \|_2)} \exp{(- \sigma_x^{-1}|x_p - x_q|) },
\end{equation}
where $\mathbf{l}_p$ is the location of pixel $p$ on the 2-D image grid, and $\sigma_l$ and $\sigma_x$ are parameters.
The value of $w_{e_{p,q}}$ is large as the correlation between $x_p$ and $x_q$ is higher.

We also introduce the incidence matrix $\mathbf{D} \in \mathbb{R}^{|\mathcal{E}| \times N_{1}N_{2}}$ for a graph $\mathcal{G}$ as follows: each entry of $\mathbf{D}$ is defined by $\mathbf{D}_{e_{p,q},k} := -1$, if $p=k$; $1$, if $q=k$; and $0$, otherwise.
Thus, $\mathbf{WD}$ is the weighted spatial difference operator defined via the graph $\mathcal{G}$.

\section{Proposed Method}
\label{sec:proposed}

\subsection{Graph Spatio-Spectral Total Variation}
\label{subseq:GSSTV}

This section is devoted to establishing GSSTV, where the spatial difference operator of SSTV is replaced by the weighted spatial difference operator defined via the graph constructed from a guide image.

Since we only have a noisy HSI in real situations, we generate the guide image by averaging all the bands of the noise HSI along the spectral direction.
Specifically, for a given HSI $\mathbf{x} \in \mathbb{R}^{N}$, the guide image $\mathbf{x}^{\prime} \in \mathbb{R}^{N_{1} N_{2}}$ is computed by
\begin{equation}
    \label{eq:guide_image}
    [\mathbf{x}^{\prime}]_{i,j} := \tfrac{1}{N_3} \sum\nolimits_{k=1}^{N_3}[\mathbf{x}]_{i,j,k}.
\end{equation}
This process yields a grayscale image that has the same spatial structure of the target HSI with noise attenuation.

Therefore, we construct a spatial graph $\mathcal{G}$ using the grayscale image as a guide image, and define a weighted spatial difference operator $\mathbf{WD}$ via this graph using the procedure described in Sec.~\ref{subsec:GTV}. Here, the number of connectable pixels is limited to eight neighborhoods to prevent the number of edges from exploding and increasing the computational cost.
The flow of the guide image generation and graph construction is shown in Fig.~\ref{fig:graph_construction}.

Finally, our GSSTV is given as follows:
\begin{equation}
	\label{eq:GSSTV}
	\text{GSSTV}_{\mathcal{G}(\mathbf{x}^{\prime},\mathcal{E},\mathbf{W})}(\mathbf{u}) := \|\mathbf{D}_{\mathcal{G}} \mathbf{D}_b \mathbf{u} \|_1,
\end{equation}
where
\begin{equation}
	\label{eq:diagWD}
	\mathbf{D}_{\mathcal{G}} = \mathrm{diag}(\overbrace{\mathbf{WD}, \dots, \mathbf{WD}}^{N_3}).
\end{equation}
The design of GSSTV is detailed in Fig.~\ref{fig:GSSTV}.

In the above definition of GSSTV, applying $\mathbf{D}_{\mathcal{G}} \mathbf{D}_b$ to an HSI data $\mathbf{u}$ is equivalent to first performing sparsification based on spectral correlation via spectral differencing, and then performing sparsification based on spatial structure via graph-based weighted spatial differencing. Thus, we can say that taking the $\ell_1$ norm of $\mathbf{D}_{\mathcal{G}} \mathbf{D}_b\mathbf{u}$ is a reasonable regularization model that exploits both the spatial and spectral structures of the target HSI. Since Eq. (5) is obviously a lower semicontinuous convex function over $\mathbb{R}^{|\mathcal{E}|N}$ (i.e. $\mathbf{D}_{\mathcal{G}} \mathbf{D}_{b} \mathbf{u}$) , we can solve it using the convex optimization technique.

\input{./fig_graph_construction}
\input{./fig_GSSTV}

\subsection{Problem Formulation}
\label{subseq:Prob_Form}
Consider that an observed hyperspectral image (of size $N_{1}\times N_{2}\times N_{3}$) $\mathbf{v}\in\mathbb{R}^{N}$ is modeled by
\begin{equation}
	\label{eq:observation_model}
	\mathbf{v} = \bar{\mathbf{u}} + \bar{\mathbf{s}} + \mathbf{n},
\end{equation}
where $\bar{\mathbf{u}}$, $\bar{\mathbf{s}}$, and $\mathbf{n}$ represent a true hyperspectral image of interest, a sparsely distributed noise such as outliers, and a random noise, respectively. Based on above observation model, the GSSTV-regularized denoising problem is formulated as a convex optimization problem with the following form:
\begin{equation}
	\label{eq:GSSTV_denoising_problem}
	\min_{\mathbf{u},\mathbf{s} \in \mathbb{R}^{N}} \text{GSSTV}_{\mathcal{G}(\mathbf{v}^{\prime},\mathcal{E},\mathbf{W})}(\mathbf{u}) \:\mathrm{s.t.}\:\begin{cases} \mathbf{u} + \mathbf{s}  \in B^{\mathbf{v}}_{2, \varepsilon}, \\  \mathbf{s}  \in B_{1, \eta}, \\ \mathbf{u} \in R_{\underline{\mu},\bar{\mu}}, \end{cases}
\end{equation}
where
\begin{align}
	\label{eq:constraint_l2ball}
	&B^{\mathbf{v}}_{2, \varepsilon} := \{ \mathbf{z}\in\mathbb{R}^{N} | \: \|\mathbf{z}-\mathbf{v}\|_2 \leq \varepsilon \},  \\
	\label{eq:constraint_l1ball}
	&B_{1, \eta} := \{ \mathbf{z}\in\mathbb{R}^{N} | \: \|\mathbf{z}\|_{1} \leq \eta \},  \\
	\label{eq:constraint_box}
	&R_{\underline{\mu},\bar{\mu}} := \{ \mathbf{z} \in \mathbb{R}^{N} | \: \underline{\mu} \leq z_i \leq \bar{\mu}  \: (i = 1,\dots , N) \}.
\end{align}
The first constraint serves as data-fidelity with the $\mathbf{v}$-centered $\ell_2$ ball with the radius $\varepsilon > 0$.
The second constraint characterizes sparse noise the zero-centered $\ell_1$ ball with the radius $\eta > 0$.
Using such a data-fidelity constraint instead of an additive data-fidelity term makes it easy to adjust hyperparameters since $\varepsilon$ and $\eta$ can be determined based only on noise intensity (independent of the other terms in the objective function), as addressed, for example, in~\cite{CSALSA,EPIpre,ono_2015,ono_2019}.
The third constraint is a box constraint with $\underline{\mu} < \bar{\mu}$ which represents the dynamic range of $\mathbf{u}$.

\subsection{Optimization}
\label{subsec:Optim}
We use a primal-dual splitting method (PDS)~\cite{PDS_pock} to solve Prob.~\eqref{eq:GSSTV_denoising_problem}. PDS can solve the following optimization problem:
\begin{equation}
	\label{eq:pds_Optim}
	\min_{\mathbf{x}\in\mathbb{R}^{K}, \mathbf{y}\in\mathbb{R}^{M}} f_1(\mathbf{x})+f_2(\mathbf{y}) \mbox{ s.t. }\mathbf{y} = \mathbf{Ax},
\end{equation}
where $f_1$ and $f_2$ are proximable convex functions\footnote{
The proximity operator of index $\gamma > 0$ of a proper lower semicontinuous convex function $f$ is defined by $\mathrm{prox}_{\gamma f}(\mathbf{x}) := \arg\min_{\mathbf{y}} f(\mathbf{y}) + \frac{1}{2\gamma} \| \mathbf{x} - \mathbf{y} \|_2^2$. If the proximity operators of $f$ is computable, we call $f$ proximable.}, $\mathbf{x} \in \mathbb{R}^{K}$ is the primal variable, representing the objective variable, $\mathbf{y} \in \mathbb{R}^{M}$ is the dual variable, serving as the auxiliary variable to aid in optimization, and $\mathbf{A} \in \mathbb{R}^{M \times K} $ is the matrix that describes the relation between the two variables.

The algorithm is given by
\begin{align}
\label{eq:pds_algo}
    &\mathbf{x}^{(t+1)} = \mathrm{prox}_{\gamma_1 f_1} [ \mathbf{x}^{(t)} - \gamma_1(\mathbf{A}^\top \mathbf{y}^{(t)}) ], \\
    &\mathbf{y}^{(t+1)} = \mathrm{prox}_{\gamma_2 f_2^{*}} [ \mathbf{y}^{(t)} - \gamma_2 \mathbf{A} \bigl(2\mathbf{x}^{(t+1)} - \mathbf{x}^{(t)} \bigr) ],
\end{align}
where $f_2^{*}$ the \textit{Fenchel--Rockafellar conjugate function}\footnote{The \textit{Fenchel--Rockafellar conjugate function} of $f$ is defined by $f^{*}(\xi) := \sup_{\mathbf{x}\in \mathbb{R}^K} \{ \langle \mathbf{x}, \mathbf{\xi} \rangle - f(\mathbf{x}) \}$. The proximity operator of $f^{*}$ can be expressed as $\mathrm{prox}_{\gamma f^{*}} (\mathbf{x}) = \mathbf{x} - \gamma \mathrm{prox}_{\gamma^{-1} f} (\gamma^{-1} \mathbf{x})$.}
of $f_2$, and stepsizes $\gamma_1,\gamma_2 > 0$ satisfy $\gamma_1 \gamma_2 \lambda_1(\mathbf{A}^\top \mathbf{A}) < 1$ ($\lambda_1(\cdot)$ stands for the maximum eigenvalue of $(\cdot)$ ). Under some mild conditions on $f_1$, $f_2$, and $\mathbf{A}$, the sequence $(\mathbf{x}^{(t)})_{t \in \mathbb{N}}$ converges to an optimal solution of Prob.~\eqref{eq:pds_Optim}.

Using the indicator functions\footnote{The indicator function of a closed convex set $C$ is defined by $\iota_C(\x):=0$, if $\x\in C$; $\infty$, otherwise.} of $ B^{\mathbf{v}}_{2, \varepsilon} $, $B_{1, \eta} $, and $R_{\underline{\mu},\bar{\mu}} $, we can rewrite Prob. \eqref{eq:GSSTV_denoising_problem} into the equivalent form:
\begin{equation}
     \label{eq:objective_indicator}
     \min_{\mathbf{u}, \mathbf{s} \in \mathbb{R}^{N}} \|\mathbf{D}_{\mathcal{G}} \mathbf{D}_b \mathbf{u}\|_1 +  \iota_{B^{\mathbf{v}}_{2, \varepsilon}}(\mathbf{u} + \mathbf{s}) + \iota_{B_{1, \eta}}(\mathbf{s}) + \iota_{R_{\underline{\mu},{\bar{\mu}}}}(\mathbf{u}).
\end{equation}

Now, let $\mathbf{x} := (\mathbf{u}^\top \: \mathbf{s}^\top)^\top$ and $\mathbf{y} := (\mathbf{y}_1^\top \: \mathbf{y}_2^\top)^{\top}$ with $\mathbf{y}_1, \mathbf{y}_2 \in \mathbb{R}^{N}$.
Then, by defining $f_1(\mathbf{x}) := \iota_{R_{\underline{\mu},{\bar{\mu}}}}(\mathbf{u}) + \iota_{B_{1, \eta}}(\mathbf{s})$, $f_2(\mathbf{y}) := \|\mathbf{y}_1\|_1 + \iota_{B^{\mathbf{v}}_{2, \varepsilon}}(\mathbf{y}_2)$, and
\begin{equation}
	\label{eq:objective2pds}
	\mathbf{A} := \begin{pmatrix} \mathbf{D}_{\mathcal{G}} \mathbf{D}_b & \mathbf{O} \\ \mathbf{I} & \mathbf{I} \end{pmatrix} ,
\end{equation}
Prob.~(\ref{eq:objective_indicator}) is reduced to Prob.~(\ref{eq:pds_Optim}), where the variables in each term are fully decoupled. This facilitates the computation of the proximity operators when solving the problem by PDS, namely, the proximity operators of $f_{1}$ and $f_{2}$ are decomposed into those of each term.

The specific calculations of each proximity operator are explained below.
The proximity operators of $\iota_{R_{\underline{\mu},{\bar{\mu}}}}$, $\iota_{B^{\mathbf{v}}_{2, \varepsilon}}$, and $\iota_{B_{1, \eta}}$ are equivalent to the projections onto each convex set. The projection onto $R_{\underline{\mu},{\bar{\mu}}}$ can be computed by pushing values outside $[\underline{\mu},{\bar{\mu}}]$ into this interval.
The projection onto $B^{\mathbf{v}}_{2, \varepsilon}$ is computed, for $\mathbf{z}\notin B^{\mathbf{v}}_{2, \varepsilon}$ by $\mathbf{v} + \frac{\varepsilon (\mathbf{z} - \mathbf{v})}{\| \mathbf{z} - \mathbf{v} \|_2}$.
The projection onto $B_{1, \eta}$ can be efficiently computed by the $\ell_1$ ball projection algorithm~\cite{L1Ball}. The proximity operator of the $\ell_1$ norm is given by the well-known soft-thresholding operation: for each entry of $\mathbf{z}\in\mathbb{R}^N$, $\mathrm{sgn}(z_i) \max \left\{ 0, |z_i| - \gamma \right\}$.

We show the detailed algorithm based on PDS in Algorithm~\ref{algo1}.
For the computational complexity, lines 2, 4, 5, 6, and 7 are $\mathcal{O}(N)$. 
This is due to the fact that the computation of the difference operators, the metric projection onto the $\ell_2$ ball, the metric projection onto the box constraint, and the soft-thresholding process are each $\mathcal{O}(N)$. Note that the computational complexity of the weighted spatial difference operator $\mathbf{D}_\mathcal{G}$ is $\mathcal{O}(N)$, since the edges of the spatial graph generated from the guide image are only stretched to neighboring pixels.
On the other hand, line 3 is $\mathcal{O}(N \log N)$ computation. 
This is due to the projection onto the $\ell_1$ ball. 
Thus, the overall complexity of one iteration of Alg.~\ref{algo1} is $\mathcal{O}(N \log N)$.

\begin{algorithm}[t]
	\caption{PDS-based algorithm for solving (\ref{eq:GSSTV_denoising_problem})}
	\label{algo1}
	\begin{algorithmic}[1]
		\renewcommand{\algorithmicrequire}{\textbf{Input:}}
		\renewcommand{\algorithmicensure}{\textbf{Output:}}
		\REQUIRE $\mathbf{u}^{(0)}, \mathbf{s}^{(0)}, \mathbf{y}_i^{(0)}(i=1,2)$
		\ENSURE $\mathbf{u}^{(t)}$
		\WHILE {A stoping criterion is not satisfied}
		\STATE $\mathbf{u}^{(t+1)} = \mathrm{prox}_{\gamma_1 \iota_{R_{\underline{\mu},{\bar{\mu}}}}} ( \mathbf{u}^{(t)} - \gamma_1 (\mathbf{D}_b^\top \mathbf{D}_{\mathcal{G}}^{\top} \mathbf{y}_1^{(t)} +  \mathbf{y}_2^{(t)} ) )$;
		\STATE $\mathbf{s}^{(t+1)} = \mathrm{prox}_{\gamma_1 \iota_{B_{1, \eta}}}( \mathbf{s}^{(t)} - \gamma_1 \mathbf{y}_2^{(t)} )$;
		\STATE $\mathbf{y}_1^{(t)} \leftarrow \mathbf{y}_1^{(t)} + \gamma_2 ( \mathbf{D}_{\mathcal{G}} \mathbf{D}_b ( 2 \mathbf{u}^{(t+1)} - \mathbf{u}^{(t)} )  )$;
		\STATE $\mathbf{y}_2^{(t)} \leftarrow \mathbf{y}_2^{(t)} +\gamma_2 ( 2 \mathbf{u}^{(t+1)} - \mathbf{u}^{(t)} + 2 \mathbf{s}^{(t+1)} - \mathbf{s}^{(t)} )$;
		\STATE $\mathbf{y}_1^{(t+1)} = \mathbf{y}_1^{(t)} - \gamma_2 \mathrm{prox}_{ \gamma_2^{-1} \| \cdot \|_1} ( \gamma_2^{-1}\mathbf{y}_1^{(t)})$;
		\STATE $\mathbf{y}_2^{(t+1)} = \mathbf{y}_2^{(t)} - \gamma_2 \mathrm{prox}_{ \gamma_2^{-1} \iota_{B^{\mathbf{v}}_{2, \varepsilon}}} ( \gamma_2^{-1} \mathbf{y}_2^{(t)} )$;
		\STATE $t \leftarrow t+1$;
		\ENDWHILE
	\end{algorithmic}
\end{algorithm}

\input{tab_results}
\input{./fig_result_image.tex}

\section{Experiments}
\label{sec:exp}

To demonstrate the effectiveness of GSSTV, we conducted mixed noise removal experiments, where we compare GSSTV with TV-based methods: Hyperspectral Total variation (HTV)~\cite{HTV}, Graph Total Variation (GTV)~\cite{couprie2013dual}, and SSTV~\cite{SSTV}; tensor-based methods: Decomposable Nonlocal Tensor Dictionary Learning (TDL)~\cite{TDL} and Intrinsic Tensor Sparsity Regularization (ITSReg)~\cite{ITSReg}; and a TV-tensor hybrid method: Total Variation-regularized Low-Rank Tensor Decomposition (LRTDTV)~\cite{LRTDTV}.
Here, GTV evaluates the $\ell_1$ norm of weighted spatial differences based on the same spatial graph as GSSTV for all bands.
We used \textit{Beltsville} from SpecTIR~\cite{Beltsville} and \textit{Pavia} from GIC ~\cite{Pavia} as ground-truth HSIs.
The Beltsville image is cropped to a size of $64 \times 64 \times 128$ and the Pavia image to $64 \times 64 \times 102$ and the pixel values are normalized to $[0,1]$.
The ground-truth images were contaminated by additive white Gaussian noise $\mathbf{n}$ and salt-and-paper sparse noise $\mathbf{s}$, where we set the standard deviation of Gaussian noise to $0.05$ or $0.1$, and the ratio of salt-and-pepper noise to $0.05$.

For a fair comparison, the TV-based methods, HTV, GTV, and SSTV, were replaced with GSSTV in \eqref{eq:GSSTV_denoising_problem}, respectively, and we solve each problem by PDS.
For TDL, ITSReg, and LRTDTV, we used the implementations published by the authors of each paper.
The parameters $\varepsilon$ in (\ref{eq:constraint_l2ball}) and $\eta$ (\ref{eq:constraint_l1ball}) were set to the oracle values in every method, i.e., $\varepsilon = \|\mathbf{n} \|_2 $ and $\eta = \|\bar{\mathbf{s}} \|_1$.
We set the stopping criteria of Alg.~\ref{algo1} to $\| \mathbf{u}^{(n+1)} - \mathbf{u}^{(n)} \|_2/\|\mathbf{u}^{(n)}\|_{2} < 1.0 \times 10^{-4}$.
The stepsizes $\gamma_1$ and $\gamma_2$ were fixed as $0.1$ and $1/(1800 \gamma_1)$ in Alg.~\ref{algo1}, which satisfy $\gamma_1 \gamma_2 \lambda_1(\mathbf{A}^\top \mathbf{A}) < 1$. 
The parameters $\sigma_l$ and $\sigma_x$ for the weights of the graph were experimentally adjusted to achieve good results.
For quality measures, we employed the mean peak signal-to-noise ratio (MPSNR) [dB]: 
$\frac{1}{N_{3}} \sum_{k=1}^{N_{3}} 10\log_{10}(N_{1} N_{2} / \| \mathbf{u}_{k} - \bar{\mathbf{u}}_{k} \|_2^2)$
, and the mean structural similarity index (MSSIM)~\cite{MSSIM}: $\frac{1}{N_{3}} \sum_{k=1}^{N_{3}} \mathrm{SSIM}(\mathbf{u}_{k}, \bar{\mathbf{u}}_{k})$, where $\mathbf{u}_{k}$ is the $k$th band of $\mathbf{u}$.
\vspace{-0.1mm}

Table~\ref{tab:MPSNR_MSSIM} shows MPSNRs [dB] and MSSIMs of all denoising results by each method.
Our GSSTV achieves better restoration than all the TV-based methods, HTV, GTV, and SSTV, and the tensor-based methods, TDL and ITSReg.
On the other hand, LRTDTV, a hybrid method combining tensor and TV, show higher performance than GSSTV in most cases.
In this regard, it should be noted that since GSSTV is a TV-based method, it could be incorporated into hybrid methods such as LRTDTV to further enhance them.
To demonstrate this, we implemented LRTDTV (LRTDTV+GSSTV), in which the SSTV term of LRTDTV was replaced by GSSTV.
The results show that LRTDTV+GSSTV has a higher denoising ability than LRTDTV in most cases, indicating that GSSTV has high potential as a TV-based regularization.

The top row of Fig.~\ref{fig:result_image} shows the denoising results at $\sigma=0.05$ and $s_p=0.05$ in Beltsville, and the bottom row those at $\sigma=0.1$ and $s_p=0.05$ in Pavia, with $[45, 75, 105]$ and $[18,59,100]$ bands visualized as color images, respectively.
The image restored by HTV in (c) shows excessive spatial smoothing and flattening of detailed structures such as edges and textures. The image restored by GTV in (d), compared to HTV, restores rough spatial structures such as edges, but all fine-scale changes are over-smoothed. These can be attributed to the fact that HTV and GTV do not properly incorporate spectral correlations.
The restored image by SSTV in (e) preserves more fine structure than the restored image by HTV or GTV, but the overall contrast is lower than the ground-truth image. This may be due to the fact that SSTV evaluates spatial differences uniformly. 
In images (f) and (g) restored with the tensor-based methods TDL and ITSReg, edges and textures are restored to some extent, but the mean pixel values are significantly off due to spectral distortion, and much noise remains in the low light bands. On the other hand, as shown in (i), GSSTV is able to recover a clean HSI from a noisy observation that preserves edges and textures. Specifically, in the Beltsville image, the black line component in the upper half of the image is maintained at the same intensity as the ground-truth, and in the Pavia image, only the noise is removed without destroying the small circular structure composed of green and red in the magnified area. The hybrid method (h), LRTDTV, is able to remove noise while preserving the detailed structure. Finally, LRTDTV+GSSTV in (j) recovers a clearer, higher quality image than any other method.

We measured the execution time [sec] per iteration of each method using MATLAB (R2022a) on a Windows 10 computer with an Intel Core i9-10900 3.7GHz processor, 32GB RAM, and an NVIDIA GeForce RTX 3090.  For HTV, GTV, SSTV, TDL, ITSReg, LRTDTV, GSSTV, and LRTDTV+GSSTV, the results are 0.0097, 0.0104, 0.0091, 0.1081, 60.0326, 0.1380, 0.0093, and 3.0086, respectively. The results show that GSSTV is comparable to other TVs in terms of execution efficiency.

We also examined the sensitivity of the edge parameters ($\sigma_l, \sigma_x$) of the spatial graph in the proposed method. Fig.~\ref{fig:parameter_sensitivity} shows the change in MPSNR and MSSIM of the restored HSI when $\sigma_l$ or $\sigma_x$ is changed. As can be seen from these plots, when $\sigma_l$ is greater than 1 and $\sigma_x$ is greater than 0.05, GSSTV can always achieve high restoration accuracy.

\input{fig_parameter_sensitivity}


\section{Conclusion}
\label{sec:Conc}
We have proposed a new HSI regularization model, named GSSTV, for HSI denoising. 
GSSTV is designed by incorporating a graph-based weighted spatial difference operator into SSTV, leading to a powerful regularization model that fully captures the spatio-spectral structure of the target HSI.
HSI denoising using GSSTV is formulated as a convex optimization problem and is efficiently solved by PDS. The experimental results on mixed noise removal illustrate the advantage of GSSTV over existing HSI denoising methods.


%





\ifCLASSOPTIONcaptionsoff
  \newpage
\fi



%




\bibliographystyle{IEEEtran}
\input{./bib_tex/bib/refs.bbl}

%








\end{document}

%% file: fig_graph_construction.tex
\begin{figure}[!t]

		\begin{minipage}{\hsize}
    	    \centerline{\includegraphics[width=\hsize]{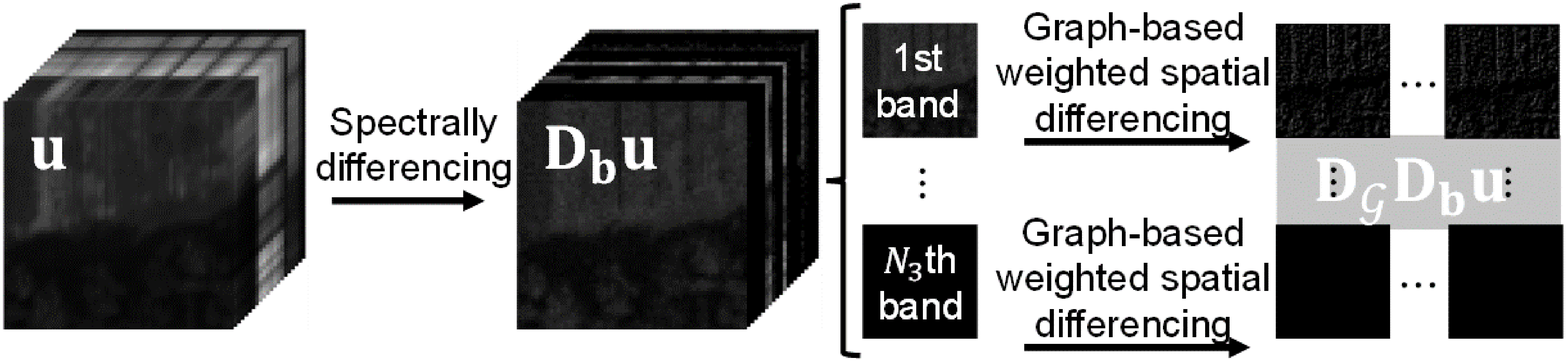}} 
    	\end{minipage}
	
	
	\caption{Guide image generation and graph construction.}
	\label{fig:graph_construction}
\end{figure}

%% file: fig_GSSTV.tex
\begin{figure}[!t]

		\begin{minipage}{\hsize}
    	    \centerline{\includegraphics[width=\hsize]{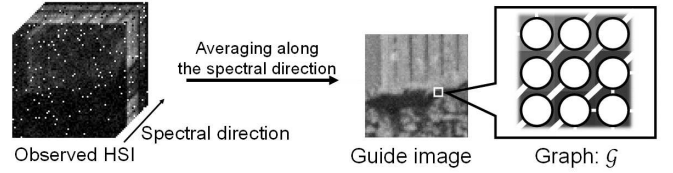}} 
    	\end{minipage}
	
	
	\caption{Design of the proposed GSSTV.}
	\label{fig:GSSTV}
\end{figure}

%% file: tab_results.tex
\begin{table*}[t]
    \begin{center}
        \caption{MPSNRs and MSSIMs of All Denoising Results.}
        \label{tab:MPSNR_MSSIM}
        		\scalebox{0.90}{
        \begin{tabular}{ccc cccccccc}
            \toprule
                Image & $(\sigma,s_p)$ & & HTV \cite{HTV} & GTV \cite{couprie2013dual} & SSTV \cite{SSTV} & TDL \cite{TDL} & ITSReg \cite{ITSReg} & LRTDTV \cite{LRTDTV} & \textbf{GSSTV} & LRTDTV+\textbf{GSSTV} \\
            \cmidrule(lr){1-11}
            \vspace{-0.5mm}
                Beltsville & (0.05, 0.05) & MPSNR & 30.41 & 29.81 & 32.69 & 21.07 & 31.71 & 38.58 & 36.05 & \textbf{39.59}  \\
            \vspace{-0.5mm}
                & & MSSIM & 0.7935 & 0.7650 & 0.8818 & 0.7854 & 0.8715 & 0.9621 & 0.9216 & \textbf{0.9649} \\
            \cmidrule(lr){2-11}
            \vspace{-0.5mm}
                & (0.1, 0.05) & MPSNR & 26.92 & 25.84 & 26.70 & 20.76 & 31.04 & \textbf{35.12} & 32.06 & 34.60 \\
            \vspace{-0.5mm}
                &  & MSSIM & 0.6084 & 0.5680 & 0.6960 & 0.7402 & 0.8525 & 0.9233 & 0.8410 & \textbf{0.9650} \\
            \cmidrule(lr){1-11}
            \vspace{-0.5mm}
                Pavia & (0.05, 0.05) & MPSNR & 27.24 & 26.97 & 31.71 & 20.55 & 29.89 & 35.43 & 36.19 & \textbf{36.97} \\
            \vspace{-0.5mm}
                & & MSSIM & 0.8198 & 0.7906 & 0.9258 & 0.8514 & 0.8969 & 0.9540 & 0.9537 & \textbf{0.9604} \\
            \cmidrule(lr){2-11}
            \vspace{-0.5mm}
                & (0.1, 0.05) & MPSNR & 24.05 & 23.48 & 26.13 & 20.32 & 29.12 & 32.12 & 31.96 & \textbf{32.90}  \\
            \vspace{-0.5mm}
                & & MSSIM & 0.6473 & 0.6006 & 0.8193 & 0.8348 & 0.8795 & 0.9195 & 0.9157 & \textbf{0.9233} \\
            \bottomrule
        \end{tabular}
        		}
    \end{center}
\end{table*}

%% file: fig_result_image.tex
\begin{figure*}[!t]
	\begin{center}
		\begin{minipage}{0.092\hsize}
			\centerline{\includegraphics[width=\hsize]{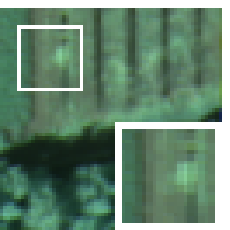}} 
		\end{minipage}
		\begin{minipage}{0.092\hsize}
			\centerline{\includegraphics[width=\hsize]{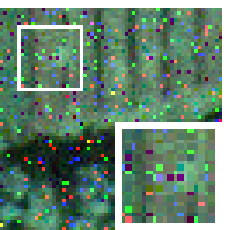}} 
		\end{minipage}
		\begin{minipage}{0.092\hsize}
			\centerline{\includegraphics[width=\hsize]{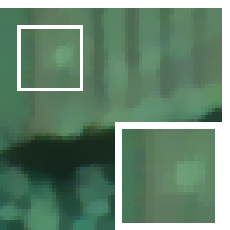}} 
		\end{minipage}
		\begin{minipage}{0.092\hsize}
			\centerline{\includegraphics[width=\hsize]{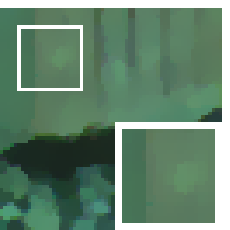}} 
		\end{minipage}
		\begin{minipage}{0.092\hsize}
			\centerline{\includegraphics[width=\hsize]{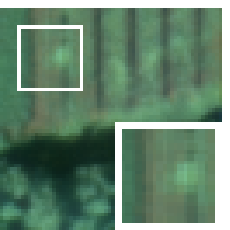}} 
		\end{minipage}
		\begin{minipage}{0.092\hsize}
			\centerline{\includegraphics[width=\hsize]{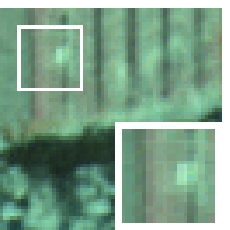}} 
		\end{minipage}
		\begin{minipage}{0.092\hsize}
			\centerline{\includegraphics[width=\hsize]{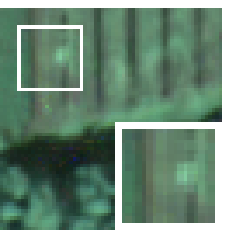}} 
		\end{minipage}
		\begin{minipage}{0.092\hsize}
			\centerline{\includegraphics[width=\hsize]{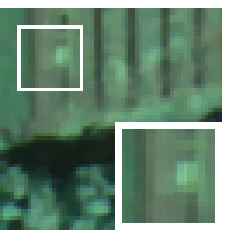}} 
		\end{minipage}
		\begin{minipage}{0.092\hsize}
			\centerline{\includegraphics[width=\hsize]{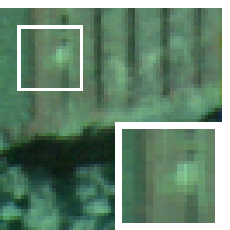}} 
		\end{minipage}
		\begin{minipage}{0.092\hsize}
			\centerline{\includegraphics[width=\hsize]{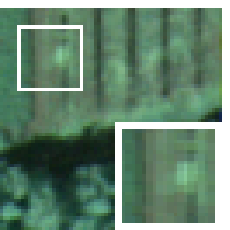}} 
		\end{minipage}
		
		\vspace{1mm}
		
		\begin{minipage}{0.092\hsize}
		    \centerline{\small{MPSNR}}
		\end{minipage}
        \begin{minipage}{0.092\hsize}
            \centerline{\small{17.06}} 
		\end{minipage}
		\begin{minipage}{0.092\hsize}
			\centerline{\small{30.41}} 
		\end{minipage}
		\begin{minipage}{0.092\hsize}
			\centerline{\small{29.81}} 
		\end{minipage}
		\begin{minipage}{0.092\hsize}
			\centerline{\small{32.69}} 
		\end{minipage}
		\begin{minipage}{0.092\hsize}
			\centerline{\small{21.07}} 
		\end{minipage}
		\begin{minipage}{0.092\hsize}
			\centerline{\small{31.71}} 
		\end{minipage}
		\begin{minipage}{0.092\hsize}
			\centerline{\small{38.58}} 
		\end{minipage}
		\begin{minipage}{0.092\hsize}
			\centerline{\small{36.05}} 
		\end{minipage}
		\begin{minipage}{0.092\hsize}
			\centerline{\textbf{\small{39.59}}} 
		\end{minipage}
		
		\begin{minipage}{0.092\hsize}
		    \centerline{\small{MSSIM}}
		\end{minipage}
        \begin{minipage}{0.092\hsize}
            \centerline{\small{0.2982}} 
		\end{minipage}
		\begin{minipage}{0.092\hsize}
			\centerline{\small{0.7935}} 
		\end{minipage}
		\begin{minipage}{0.092\hsize}
			\centerline{\small{0.7650}} 
		\end{minipage}
		\begin{minipage}{0.092\hsize}
			\centerline{\small{0.8818}} 
		\end{minipage}
		\begin{minipage}{0.092\hsize}
			\centerline{\small{0.7854}} 
		\end{minipage}
		\begin{minipage}{0.092\hsize}
			\centerline{\small{0.8715}} 
		\end{minipage}
		\begin{minipage}{0.092\hsize}
			\centerline{\small{0.9621}} 
		\end{minipage}
		\begin{minipage}{0.092\hsize}
			\centerline{\small{0.9216}} 
		\end{minipage}
		\begin{minipage}{0.092\hsize}
			\centerline{\textbf{\small{0.9649}}} 
		\end{minipage}
		
		\vspace{1mm}
		
		\begin{minipage}{0.092\hsize}
			\centerline{\includegraphics[width=\hsize]{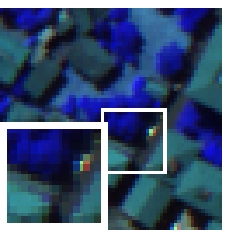}} 
		\end{minipage}
		\begin{minipage}{0.092\hsize}
			\centerline{\includegraphics[width=\hsize]{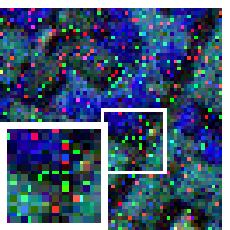}} 
		\end{minipage}
		\begin{minipage}{0.092\hsize}
			\centerline{\includegraphics[width=\hsize]{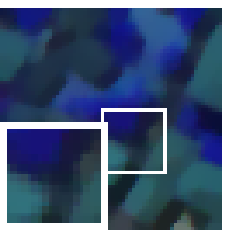}} 
		\end{minipage}
		\begin{minipage}{0.092\hsize}
			\centerline{\includegraphics[width=\hsize]{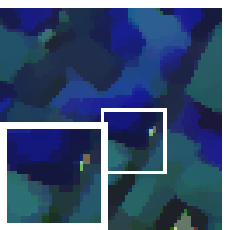}} 
		\end{minipage}
		\begin{minipage}{0.092\hsize}
			\centerline{\includegraphics[width=\hsize]{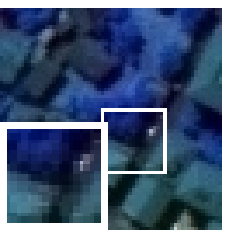}} 
		\end{minipage}
		\begin{minipage}{0.092\hsize}
			\centerline{\includegraphics[width=\hsize]{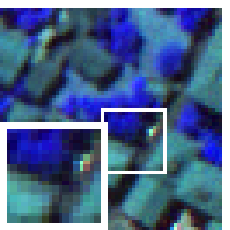}} 
		\end{minipage}
		\begin{minipage}{0.092\hsize}
			\centerline{\includegraphics[width=\hsize]{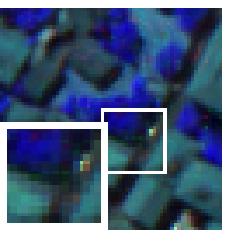}} 
		\end{minipage}
		\begin{minipage}{0.092\hsize}
			\centerline{\includegraphics[width=\hsize]{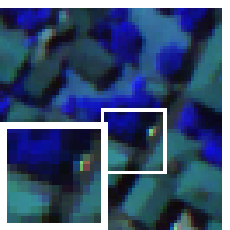}} 
		\end{minipage}
		\begin{minipage}{0.092\hsize}
			\centerline{\includegraphics[width=\hsize]{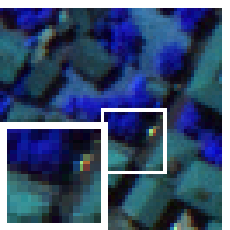}} 
		\end{minipage}
		\begin{minipage}{0.092\hsize}
			\centerline{\includegraphics[width=\hsize]{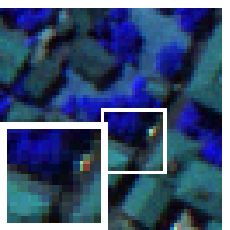}} 
		\end{minipage}
		
		\vspace{1mm}
		
		\begin{minipage}{0.092\hsize}
		    \centerline{\small{MPSNR}}
		\end{minipage}
        \begin{minipage}{0.092\hsize}
            \centerline{\small{15.78}} 
		\end{minipage}
		\begin{minipage}{0.092\hsize}
			\centerline{\small{24.05}} 
		\end{minipage}
		\begin{minipage}{0.092\hsize}
			\centerline{\small{23.48}} 
		\end{minipage}
		\begin{minipage}{0.092\hsize}
			\centerline{\small{26.13}} 
		\end{minipage}
		\begin{minipage}{0.092\hsize}
			\centerline{\small{20.32}} 
		\end{minipage}
		\begin{minipage}{0.092\hsize}
			\centerline{\small{29.12}} 
		\end{minipage}
		\begin{minipage}{0.092\hsize}
			\centerline{\small{32.12}} 
		\end{minipage}
		\begin{minipage}{0.092\hsize}
			\centerline{\small{31.96}} 
		\end{minipage}
		\begin{minipage}{0.092\hsize}
			\centerline{\textbf{\small{32.90}}} 
		\end{minipage}

		\begin{minipage}{0.092\hsize}
		    \centerline{\small{MSSIM}}
		\end{minipage}
        \begin{minipage}{0.092\hsize}
            \centerline{\small{0.3627}} 
		\end{minipage}
		\begin{minipage}{0.092\hsize}
			\centerline{\small{0.6473}} 
		\end{minipage}
		\begin{minipage}{0.092\hsize}
			\centerline{\small{0.6006}} 
		\end{minipage}
		\begin{minipage}{0.092\hsize}
			\centerline{\small{0.8193}} 
		\end{minipage}
	    \begin{minipage}{0.092\hsize}
			\centerline{\small{0.8348}} 
		\end{minipage}
		\begin{minipage}{0.092\hsize}
			\centerline{\small{0.8795}} 
		\end{minipage}
		\begin{minipage}{0.092\hsize}
			\centerline{\small{0.9195}} 
		\end{minipage}
		\begin{minipage}{0.092\hsize}
			\centerline{\small{0.9157}} 
		\end{minipage}
		\begin{minipage}{0.092\hsize}
			\centerline{\textbf{\small{0.9233}}} 
		\end{minipage}

		\begin{minipage}{0.092\hsize}
			\centerline{\small{(a)}}
		\end{minipage}
		\begin{minipage}{0.092\hsize}
			\centerline{\small{(b)}}
		\end{minipage}
		\begin{minipage}{0.092\hsize}
			\centerline{\small{(c)}}
		\end{minipage}
		\begin{minipage}{0.092\hsize}
			\centerline{\small{(d)}}
		\end{minipage}
		\begin{minipage}{0.092\hsize}
			\centerline{\small{(e)}}
		\end{minipage}
		\begin{minipage}{0.092\hsize}
			\centerline{\small{(f)}}
		\end{minipage}
		\begin{minipage}{0.092\hsize}
			\centerline{\small{(g)}}
		\end{minipage}
		\begin{minipage}{0.092\hsize}
			\centerline{\small{(h)}}
		\end{minipage}
		\begin{minipage}{0.092\hsize}
			\centerline{\small{(i)}}
		\end{minipage}
		\begin{minipage}{0.092\hsize}
			\centerline{\small{(j)}}
		\end{minipage}
	\end{center}
	
	\vspace{-3mm}
	\caption{Denoising results. The top is the results on Beltsville with $\sigma=0.05$, $s_p=0.05$ and the bottom is the results on Pavia with $\sigma=0.1$, $s_p = 0.05$. (a) Ground-truth. (b) Observed noisy image. (c) HTV. (d) GTV. (e) SSTV. (f) TDL. (g) ITSReg. (h) LRTDTV. (i) GSSTV (Ours). (j) LRTDTV+GSSTV}
	\label{fig:result_image}
\end{figure*}

%% file: fig_parameter_sensitivity.tex
\begin{figure}[!t]
        \centering

		\begin{minipage}{0.24\hsize}
    	    \centerline{\includegraphics[width=\hsize]{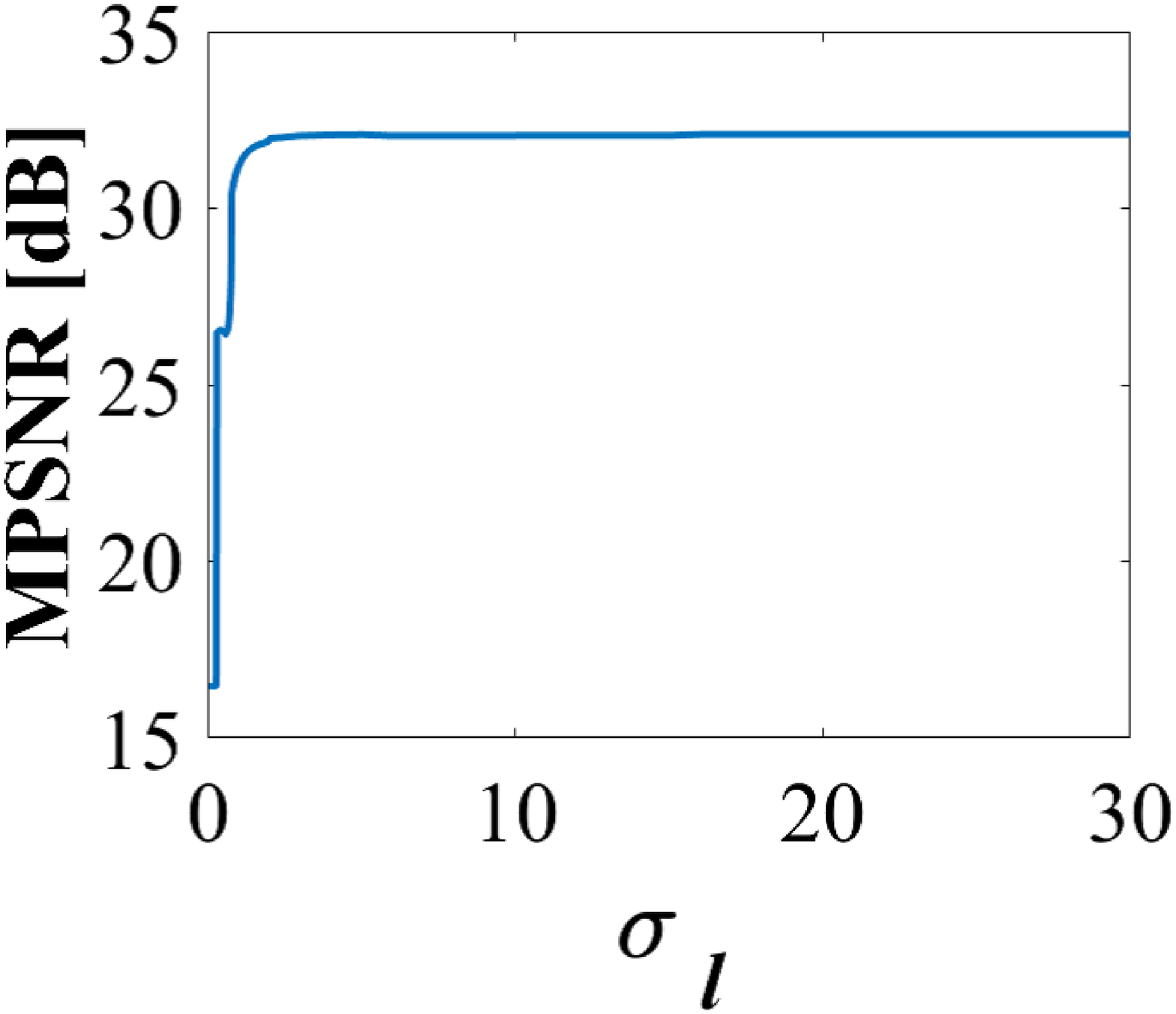}} 
    	\end{minipage}
    	\begin{minipage}{0.24\hsize}
    	    \centerline{\includegraphics[width=\hsize]{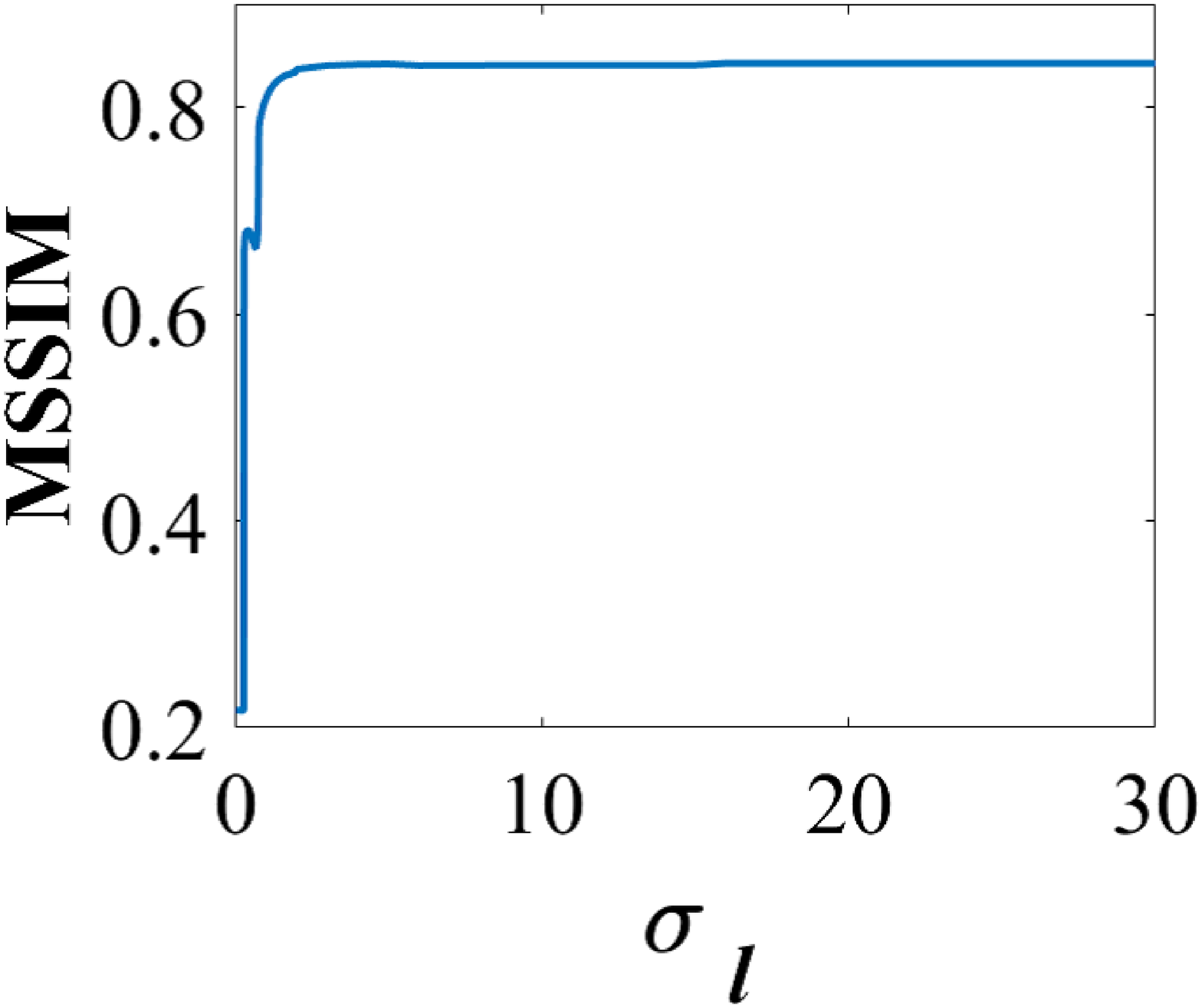}} 
    	\end{minipage}
    	\begin{minipage}{0.24\hsize}
    	    \centerline{\includegraphics[width=\hsize]{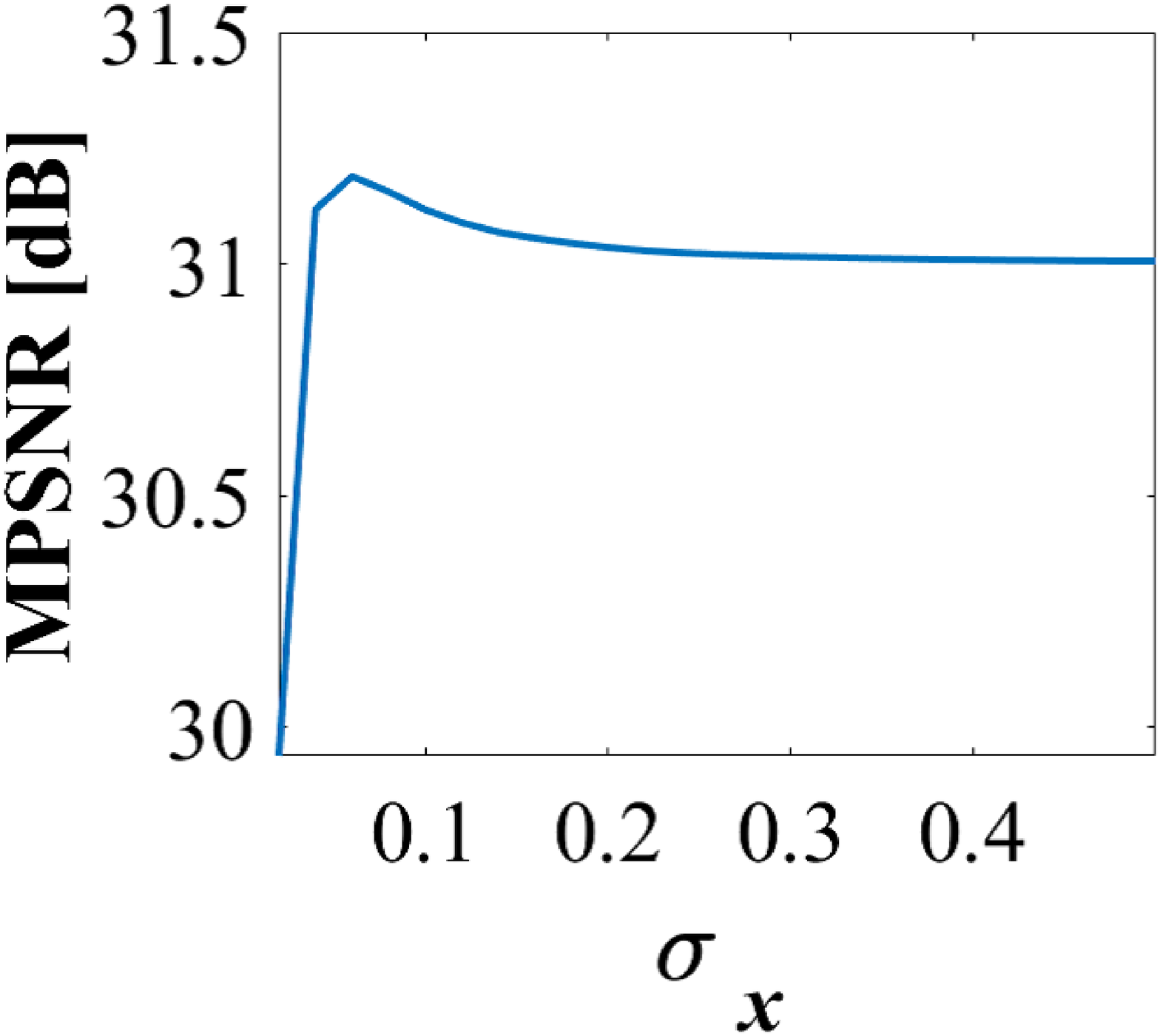}} 
    	\end{minipage}
    	\begin{minipage}{0.24\hsize}
    	    \centerline{\includegraphics[width=\hsize]{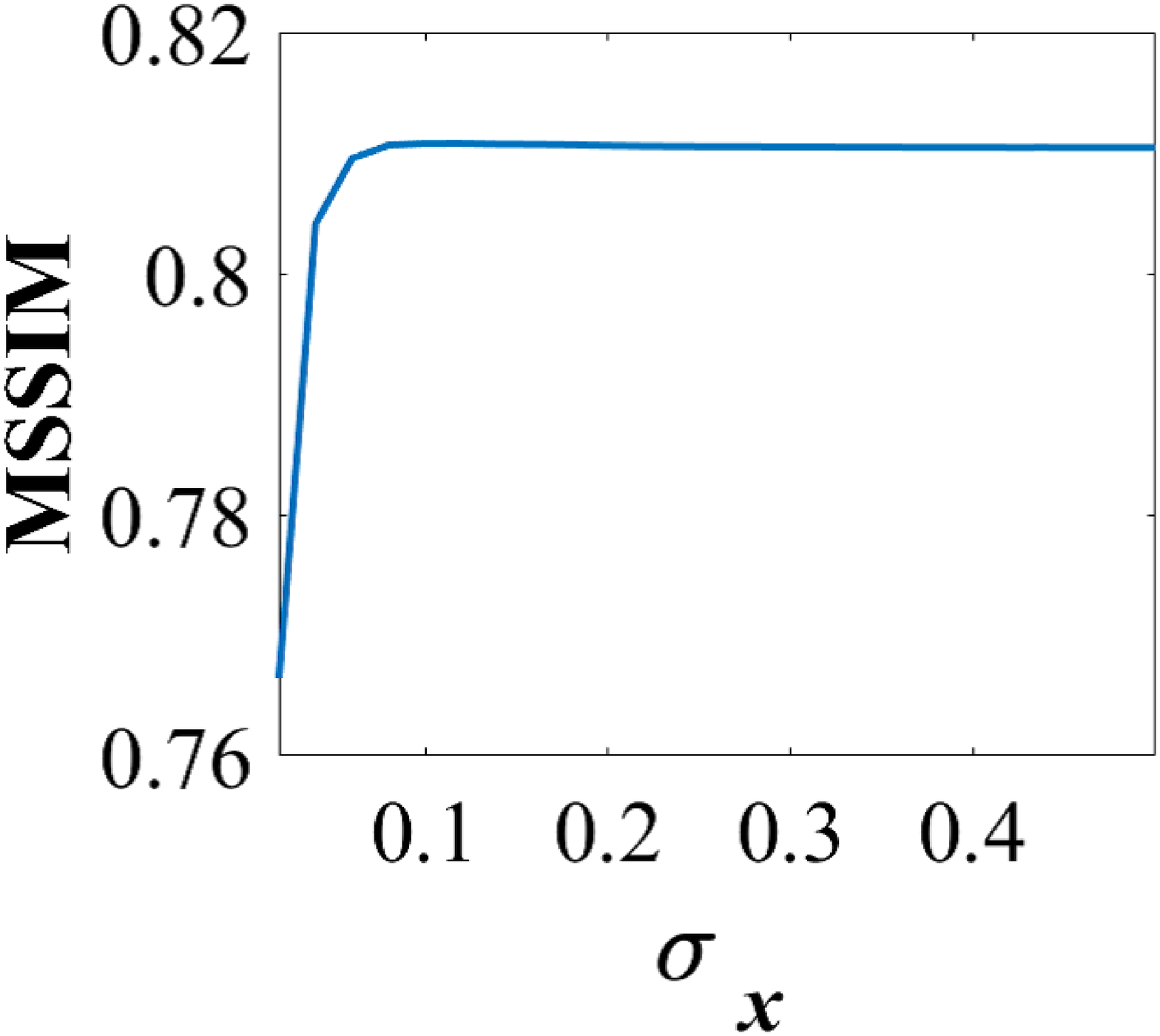}} 
    	\end{minipage}
    	
    	\begin{minipage}{0.48\hsize}
			\centerline{(a)}
		\end{minipage}
		\begin{minipage}{0.48\hsize}
			\centerline{(b)}
		\end{minipage}
	
	
	\caption{(a) MPSNR or MSSIM versus $\sigma_l$. (b) MPSNR or MSSIM versus $\sigma_x$.}
	\label{fig:parameter_sensitivity}
\end{figure}



	
	

%% file: bib_tex/bib/refs.bbl

%% file: GRSL_GSSTV_arXiv.bbl
\begin{thebibliography}{10}
\providecommand{\url}[1]{#1}
\csname url@samestyle\endcsname
\providecommand{\newblock}{\relax}
\providecommand{\bibinfo}[2]{#2}
\providecommand{\BIBentrySTDinterwordspacing}{\spaceskip=0pt\relax}
\providecommand{\BIBentryALTinterwordstretchfactor}{4}
\providecommand{\BIBentryALTinterwordspacing}{\spaceskip=\fontdimen2\font plus
\BIBentryALTinterwordstretchfactor\fontdimen3\font minus
  \fontdimen4\font\relax}
\providecommand{\BIBforeignlanguage}[2]{{%
\expandafter\ifx\csname l@#1\endcsname\relax
\typeout{** WARNING: IEEEtran.bst: No hyphenation pattern has been}%
\typeout{** loaded for the language `#1'. Using the pattern for}%
\typeout{** the default language instead.}%
\else
\language=\csname l@#1\endcsname
\fi
#2}}
\providecommand{\BIBdecl}{\relax}
\BIBdecl

\bibitem{borengasser2007hyperspectral}
M.~Borengasser, W.~S. Hungate, and R.~Watkins, \emph{Hyperspectral remote
  sensing: principles and applications}.\hskip 1em plus 0.5em minus 0.4em\relax
  CRC press, 2007.

\bibitem{grahn2007techniques}
H.~Grahn and P.~Geladi, \emph{Techniques and applications of hyperspectral
  image analysis}.\hskip 1em plus 0.5em minus 0.4em\relax John Wiley \& Sons,
  2007.

\bibitem{thenkabail2016hyperspectral}
P.~S. Thenkabail and J.~G. Lyon, \emph{Hyperspectral remote sensing of
  vegetation}.\hskip 1em plus 0.5em minus 0.4em\relax CRC press, 2016.

\bibitem{HSI_unmixing_review}
J.~M. {Bioucas-Dias}, A.~{Plaza}, N.~{Dobigeon}, M.~{Parente}, Q.~{Du},
  P.~{Gader}, and J.~{Chanussot}, ``Hyperspectral unmixing overview:
  Geometrical, stastical, and sparse regression-based approaches,'' \emph{IEEE
  J. Sel. Topics Appl. Earth Observ. Remote Sens.}, vol.~5, no.~2, pp.
  354--379, Apr. 2012.

\bibitem{unmixing2014}
W.~{Ma}, J.~M. {Bioucas-Dias}, T.~{Chan}, N.~{Gillis}, P.~{Gader}, A.~J.
  {Plaza}, A.~{Ambikapathi}, and C.~{Chi}, ``A signal processing perspective on
  hyperspectral unmixing: Insights from remote sensing,'' \emph{IEEE Signal
  Process. Mag.}, vol.~31, no.~1, pp. 67--81, Dec. 2014.

\bibitem{ghamisi2017advanced}
P.~Ghamisi, J.~Plaza, Y.~Chen, J.~Li, and A.~J. Plaza, ``Advanced spectral
  classifiers for hyperspectral images: A review,'' \emph{IEEE Geosci. Remote
  Sens. Mag.}, vol.~5, no.~1, pp. 8--32, 2017.

\bibitem{audebert2019deep}
N.~Audebert, B.~Le~Saux, and S.~Lef{\`e}vre, ``Deep learning for classification
  of hyperspectral data: A comparative review,'' \emph{IEEE Geosci. Remote
  Sens. Mag.}, vol.~7, no.~2, pp. 159--173, 2019.

\bibitem{Classification_DR}
F.~Luo, Z.~Zou, J.~Liu, and Z.~Lin, ``Dimensionality reduction and
  classification of hyperspectral image via multistructure unified
  discriminative embedding,'' \emph{IEEE Trans. Geosci. Remote Sens.}, vol.~60,
  pp. 1--16, 2022.

\bibitem{SAHDA}
F.~Luo, L.~Zhang, X.~Zhou, T.~Guo, Y.~Cheng, and T.~Yin, ``Sparse-adaptive
  hypergraph discriminant analysis for hyperspectral image classification,''
  \emph{IEEE Geosci. Remote Sens. Lett.}, vol.~17, no.~6, pp. 1082--1086, 2020.

\bibitem{rasti2018noise}
B.~Rasti, P.~Scheunders, P.~Ghamisi, G.~Licciardi, and J.~Chanussot, ``Noise
  reduction in hyperspectral imagery: Overview and application,'' \emph{Remote
  Sens.}, vol.~10, no.~3, p. 482, 2018.

\bibitem{HSI_review_Ghamisi2017}
P.~Ghamisi, N.~Yokoya, J.~Li, W.~Liao, S.~Liu, J.~Plaza, B.~Rasti, and
  A.~Plaza, ``Advances in hyperspectral image and signal processing: A
  comprehensive overview of the state of the art,'' \emph{IEEE Geosci. Remote
  Sens. Mag.}, vol.~5, no.~4, pp. 37--78, Dec. 2017.

\bibitem{SSTV}
H.~K. {Aggarwal} and A.~{Majumdar}, ``Hyperspectral image denoising using
  spatio-spectral total variation,'' \emph{IEEE Geosci. Remote Sens. Lett.},
  vol.~13, no.~3, pp. 442--446, Feb. 2016.

\bibitem{LRMRSSTV}
W.~{He}, H.~{Zhang}, H.~{Shen}, and L.~{Zhang}, ``Hyperspectral image denoising
  using local low-rank matrix recovery and global spatial-spectral total
  variation,'' \emph{IEEE J. Sel. Topics Appl. Earth Observ. Remote Sens.},
  vol.~11, no.~3, pp. 713--729, Mar. 2018.

\bibitem{SSTV_LRTD}
H.~Fan, C.~Li, Y.~Guo, G.~Kuang, and J.~Ma, ``Spatial-spectral total variation
  regularized low-rank tensor decomposition for hyperspectral image
  denoising,'' \emph{IEEE Trans. Geosci. Remote Sens.}, vol.~56, no.~10, pp.
  6196--6213, Oct. 2018.

\bibitem{GLSSTV}
T.~{Ince}, ``Hyperspectral image denoising using group low-rank and
  spatial-spectral total variation,'' \emph{IEEE Access}, vol.~7, pp.
  52\,095--52\,109, Apr. 2019.

\bibitem{SSTV_Wang2021}
M.~Wang, Q.~Wang, J.~Chanussot, and D.~Hong, ``$l_{0}$-$l_{1}$ hybrid total
  variation regularization and its applications on hyperspectral image mixed
  noise removal and compressed sensing,'' \emph{IEEE Trans. Geosci. Remote
  Sens.}, vol.~59, no.~9, pp. 7695--7710, Sep. 2021.

\bibitem{graph_image_denoising_2008}
A.~Elmoataz, O.~Lezoray, and S.~Bougleux, ``Nonlocal discrete regularization on
  weighted graphs: A framework for image and manifold processing,'' \emph{IEEE
  Trans. Image Process.}, vol.~17, no.~7, pp. 1047--1060, Jul. 2008.

\bibitem{couprie2013dual}
C.~Couprie, L.~Grady, L.~Najman, J.-C. Pesquet, and H.~Talbot, ``Dual
  constrained tv-based regularization on graphs,'' \emph{SIAM J. Imag. Sci.},
  vol.~6, no.~3, pp. 1246--1273, 2013.

\bibitem{ono2015total}
S.~Ono, I.~Yamada, and I.~Kumazawa, ``Total generalized variation for graph
  signals,'' in \emph{2015 IEEE International Conference on Acoustics, Speech
  and Signal Processing (ICASSP)}, 2015, pp. 5456--5460.

\bibitem{chen2015signal}
S.~Chen, A.~Sandryhaila, J.~M. Moura, and J.~Kova{\v{c}}evi{\'c}, ``Signal
  recovery on graphs: Variation minimization,'' \emph{IEEE Trans. Signal
  Process.}, vol.~63, no.~17, pp. 4609--4624, 2015.

\bibitem{berger2017graph}
P.~Berger, G.~Hannak, and G.~Matz, ``Graph signal recovery via primal-dual
  algorithms for total variation minimization,'' \emph{IEEE J. Sel. Topics
  Signal Process.}, vol.~11, no.~6, pp. 842--855, 2017.

\bibitem{graph_image_denoising_Gene2017}
J.~Pang and G.~Cheung, ``Graph laplacian regularization for image denoising:
  Analysis in the continuous domain,'' \emph{IEEE Trans. Image Process.},
  vol.~26, no.~4, pp. 1770--1785, Apr. 2017.

\bibitem{PDS_pock}
A.~Chambolle and T.~Pock, ``A first-order primal-dual algorithm for convex
  problems with applications to imaging,'' \emph{J. Math. Imag. Vis.}, vol.~40,
  no.~1, pp. 120--145, 2010.

\bibitem{GSIP}
G.~Cheung, E.~Magli, Y.~Tanaka, and M.~K. Ng, ``Graph spectral image
  processing,'' \emph{Proc. IEEE}, vol. 106, no.~5, pp. 907--930, 2018.

\bibitem{CSALSA}
M.~Afonso, J.~Bioucas-Dias, and M.~Figueiredo, ``An augmented {L}agrangian
  approach to the constrained optimization formulation of imaging inverse
  problems,'' \emph{IEEE Trans. Image Process.}, vol.~20, no.~3, pp. 681--695,
  Mar. 2011.

\bibitem{EPIpre}
G.~Chierchia, N.~Pustelnik, J.-C. Pesquet, and B.~Pesquet-Popescu,
  ``Epigraphical projection and proximal tools for solving constrained convex
  optimization problems,'' \emph{Signal, Image Video Process.}, vol.~9, no.~8,
  pp. 1737--1749, 2015.

\bibitem{ono_2015}
S.~Ono and I.~Yamada, ``Signal recovery with certain involved convex
  data-fidelity constraints,'' \emph{IEEE Trans. Signal Process.}, vol.~63,
  no.~22, pp. 6149--6163, Nov. 2015.

\bibitem{ono_2019}
S.~Ono, ``Efficient constrained signal reconstruction by randomized
  epigraphical projection,'' in \emph{Proc. IEEE Int. Conf. Acoust., Speech,
  Signal Process., (ICASSP)}.\hskip 1em plus 0.5em minus 0.4em\relax IEEE,
  2019, pp. 4993--4997.

\bibitem{L1Ball}
L.~Condat, ``Fast projection onto the simplex and the l1 ball,'' \emph{Math.
  Program.}, vol. 158, no.~1, pp. 575--585, Jul. 2016.

\bibitem{HTV}
Q.~Yuan, L.~Zhang, and H.~Shen, ``Hyperspectral image denoising employing a
  spectral–spatial adaptive total variation model,'' \emph{IEEE Trans.
  Geosci. Remote Sens.}, vol.~50, no.~10, pp. 3660--3677, 2012.

\bibitem{TDL}
Y.~Peng, D.~Meng, Z.~Xu, C.~Gao, Y.~Yang, and B.~Zhang, ``Decomposable nonlocal
  tensor dictionary learning for multispectral image denoising,'' in
  \emph{Proc. IEEE Conf. Comput. Vis. Pattern Recognit. (CVPR)}, 2014, pp.
  2949--2956.

\bibitem{ITSReg}
Q.~Xie, Q.~Zhao, D.~Meng, Z.~Xu, S.~Gu, W.~Zuo, and L.~Zhang, ``Multispectral
  images denoising by intrinsic tensor sparsity regularization,'' in
  \emph{Proc. IEEE Conf. Comput. Vis. Pattern Recognit. (CVPR)}, 2016, pp.
  1692--1700.

\bibitem{LRTDTV}
Y.~Wang, J.~Peng, Q.~Zhao, Y.~Leung, X.-L. Zhao, and D.~Meng, ``Hyperspectral
  image restoration via total variation regularized low-rank tensor
  decomposition,'' \emph{IEEE J. Sel. Topics Appl. Earth Observ. Remote Sens.},
  vol.~11, no.~4, pp. 1227--1243, 2018.

\bibitem{Beltsville}
\BIBentryALTinterwordspacing
``{SpecTIR}.'' [Online]. Available:
  \url{https://www.spectir.com/contact#free-data-samples.}
\BIBentrySTDinterwordspacing

\bibitem{Pavia}
\BIBentryALTinterwordspacing
``{GIC}.'' [Online]. Available:
  \url{http://www.ehu.eus/ccwintco/index.php/Hyperspectral_Remote_Sensing_Scenes#Pavia_Centre_scene.}
\BIBentrySTDinterwordspacing

\bibitem{MSSIM}
{Z. Wang}, A.~C. {Bovik}, H.~R. {Sheikh}, and E.~P. {Simoncelli}, ``Image
  quality assessment: from error visibility to structural similarity,''
  \emph{IEEE Trans. Image Process.}, vol.~13, no.~4, pp. 600--612, Apr. 2004.

\end{thebibliography}
